\documentclass[final,letterpaper,times,12pt]{elsarticle-gps}

\usepackage{graphicx}
\usepackage{amssymb}
\usepackage{color}
\usepackage{lineno}

\def\ls{\mathrel{\hbox{\rlap{\hbox{\lower4pt\hbox{$\sim$}}}\hbox{$<$}}}}
\def\gs{\mathrel{\hbox{\rlap{\hbox{\lower4pt\hbox{$\sim$}}}\hbox{$>$}}}}

\notenumber{1}
\noteversion{1.1}
\notedate{November 17, 2021}
\begin{document}

\begin{frontmatter}

\title{Strong Lensing Science Collaboration input to the on-sky commissioning of the Vera Rubin Observatory}

\author[1]{Graham P. Smith}
\author[2,3]{Timo Anguita}
\author[4]{Simon Birrer}
\author[5]{Paul L.\ Schechter}
\author[6]{Aprajita Verma}
\author[7]{Tom Collett}
\author[8]{Frederic Courbin}
\author[9]{Brenda Frye}
\author[10]{Raphael Gavazzi}
\author[8]{Cameron Lemon}
\author[11]{Anupreeta More}
\author[1]{Dan Ryczanowski}
\author[12]{Sherry H.\ Suyu}
\author{on behalf of the Strong Lensing Science Collaboration}

\address[1]{School of Physics and Astronomy, University of Birmingham, Edgbaston, B15 2TT, U.K. Email:gps@star.sr.bham.ac.uk}
\address[2]{Departamento de Ciencias Fisicas, Universidad Andres Bello Fernandez Concha 700, Las Condes, Santiago, Chile}
\address[3]{Millennium Institute of Astrophysics, Nuncio Monse\~nor Sotero Sanz 100, Providencia, Santiago 9500011, Chile}
\address[4]{Kavli Institute for Particle Astrophysics and Cosmology, Stanford University, Stanford, CA 94305, USA}
\address[5]{MIT Kavli Institute for Astrophysics and Space Research, 77 Massachusetts Avenue, Cambridge, MA 02139, USA}
\address[6]{Sub-department of Astrophysics, University of Oxford, Denys Wilkinson Building, Oxford, OX1 3RH, U.K.}
\address[7]{Institute of Cosmology and Gravitation, University of Portsmouth, Portsmouth, PO1 3FX, U.K.}
\address[8]{Institute of Physics, Laboratory of Astrophysics, 
\'Ecole Polytechnique F\'ed\'erale de Lausanne, Observatoire de Sauverny, CH-1290 Versoix, Switzerland}
\address[9]{Department of Astronomy and Steward Observatory, University of Arizona, Tucson, AZ 85721, USA}
\address[10]{CNRS and Sorbonne Universit\'e, UMR 7095, Institut d'Astrophysique de Paris, Paris, France}
\address[11]{The Inter-University Centre for Astronomy and Astrophysics, Post Bag 4, Ganeshkhind, Pune, 411007, India}
\address[12]{Max-Planck-Institut f\"ur Astrophysik, Karl-Schwarzschild-Str.\ 1, 85748, Garching, Germany}

\begin{abstract}
  We present the Strong Lensing Science Collaboration's (SLSC)
  recommended observing targets for the science verification and
  science validation phases of commissioning. Our recommendations have
  been developed in collaboration with the Dark Energy Science
  Collaboration (DESC) Strong Lensing Topical Team. In summary, our
  key recommendations are as follows:
  \begin{enumerate}
  \item Prioritize fields that span the full range of declination
    observable from Cerro Pach\'on during the engineering focused
    Science Verification phase of commissioning, before concentrating
    on equatorial fields for the Science Validation surveys.
  \item Observe quadruply lensed quasars as the ultimate test of the
    Active Optics system towards the end of the Science Verification
    phase of commissioning. These systems are the strongest tests
    known for delivered image quality (DIQ; Figure~\ref{fig:quasars}).
  \item Prioritize candidate science validation survey fields (both
    single deep pointings and wide fields) that have been searched
    thoroughly by precursor surveys for strong lenses.
  \item The optimal wide (${\simeq}100\,\rm degree^2$) science
    validation field would include the CFHT-LS W4 field, and overlap
    with the SDSS Stripe 82, DES-SN, KIDS and HSC-SSP fields.
  \item The optimal single pointing science validation fields are the
    XMM-LSS and COSMOS Deep Drilling Fields, the equatorial Hubble
    Frontier Fields galaxy clusters, and strongly lensed quasars with
    measured time delays that are well-matched to commissioning
    timescales.
\end{enumerate}

\end{abstract}

\end{frontmatter}

\section{Introduction}

\noindent
This document is the Strong Lensing Science Collaboration's (SLSC)
``Commissioning Note'', and has been prepared in response to the Rubin
Project's call for community input to the on-sky observing strategy
during
commissioning\footnote{https://community.lsst.org/t/community-input-to-the-on-sky-observing-strategy-during-commissioning/4406}. Recommendations
from the SLSC contained in this Note span both the science
verification and science validation phases of commissioning. Our
understanding is that the Rubin Project Team define the former as
primarily an engineering-focused phase in which opportunistic on-sky
observations with ComCam and LSSTCam will be performed. We understand
that science verification will be followed by sustained on-sky
observing which aims to achieve at least ten-year depth and to enable
testing of transient detection via two complementary science
validation surveys.

Strong gravitational lensing is broadly accepted as having huge
potential for major scientific breakthroughs in the next $10+$
years. These breakthroughs are expected to span a diverse range of
high impact topics including testing cold dark matter predictions on
sub-galaxy scales, measurements of the Hubble parameter independent of
the cosmic distance ladder based on time delay cosmography (e.g. Suyu
et al. 2013), novel constraints on the dark energy equation of state
parameter from multiple source-plane lenses (Jullo et al. 2010;
Collett \& Auger 2014), discovery of optical counterparts to
gravitationally lensed gravitational waves (Smith et al. 2019), and a
new window on high redshift galaxy evolution.

Our ability to achieve breakthrough strong lensing science is
fundamentally limited by the number of known strong lenses across the
dark matter halo mass function, from galaxy-scale lenses through to
the most massive cluster-scale lenses. Rubin's Wide Fast Deep survey
therefore has the potential to revolutionize strong lensing science by
expanding the number of lenses by three orders of magnitude from
hundreds to hundreds of thousands. Realising this expansion relies in
turn on the quality of the data products delivered by Rubin, both in
low latency as ``Prompt Products'' (PP) and on longer timescales as
``Data Release Products'' (DRPs) and the template images used in
difference image analysis (DIA).

In this Note we recommend commissioning fields that will enable a
step-by-step approach to helping the Rubin Construction Project pass
the Operations Readiness Review (ORR), and ensuring that Rubin data
will enable strong lensing science promptly once survey operations
commence. Each step will be critical, beginning with verifying the
performance of the active optics system with quadruply lensed quasars,
validating the quality of PPs, DRPs and template images during the
science validation phase, and then running the lens discovery pipeline
that we are developing jointly with DESC on the second data preview
(DP2). Here we describe the selection of the fields themselves, and
discuss them in relation to pre-ORR science validation and
verification tasks. Our early science plans for DP2 lie beyond the ORR
and are therefore outside the scope of this Note.

We organise our recommendations under science verification
(Section~\ref{sec:verify}) and science validation
(Section~\ref{sec:validate}), list the commissioning fields that we
recommend in Table~1, and summarize our preferred
science validation survey scenarios by quarter in Table~2.

\section{Recommended fields for Science Verification}
\label{sec:verify}

\noindent
Rubin’s delivered image quality (DIQ) is mission critical for many
high profile science goals across LSST Science Collaborations. It is
therefore essential to evaluate and optimize Rubin’s DIQ prior to the
beginning of science operations using well-defined and challenging
scientific ``use cases''. Strongly lensed point sources (i.e.\ quasars
and explosive transients) in early type galaxies are the strongest use
cases that we have identified so far, driven by the challenges of
source crowding and the resulting complex background.

The numerous strongly lensed quasars that have been discovered to date
therefore provide a valuable resource from which to select science
verification targets. In particular, lensed quasars span a wide range
of right ascension, and the brightness of the quasar images are
typically well-matched to Rubin's single-visit depth. Importantly,
lensed quasars have for many years been a classic filler/snapshot
target list from which an object can be selected quickly when
engineering allows on-sky observations.

We have selected eight quadruply lensed quasars that have multiple
image separations in the range $0.6-1.5$ arcseconds, and are thus
resolvable in a range of observing conditions
(Table~1). The targets span a range of right ascension
and declination such that they span the full calendar year and both
northern and southern skies as seen from Cerro Pach\'on. All eight
targets are at relatively low airmass for many hours each night; the
most northerly system (SDSS\,J0924$+$0219) is observable for 2--5
hours per night at $\rm airmass{<}1.5$ during November to May. Whilst
all targets have been selected for their role in testing the DIQ in
commissioning, SDSS\,J0924 is also ripe for large caustic crossing
events for which a 600\,ksec target of opportunity programme with
\emph{Chandra} has recently been awarded. This adds considerable
scientific interest to these engineering-focused observations.

\begin{figure}[ht]
  \centering
  \includegraphics[width=0.8\hsize,angle=0]{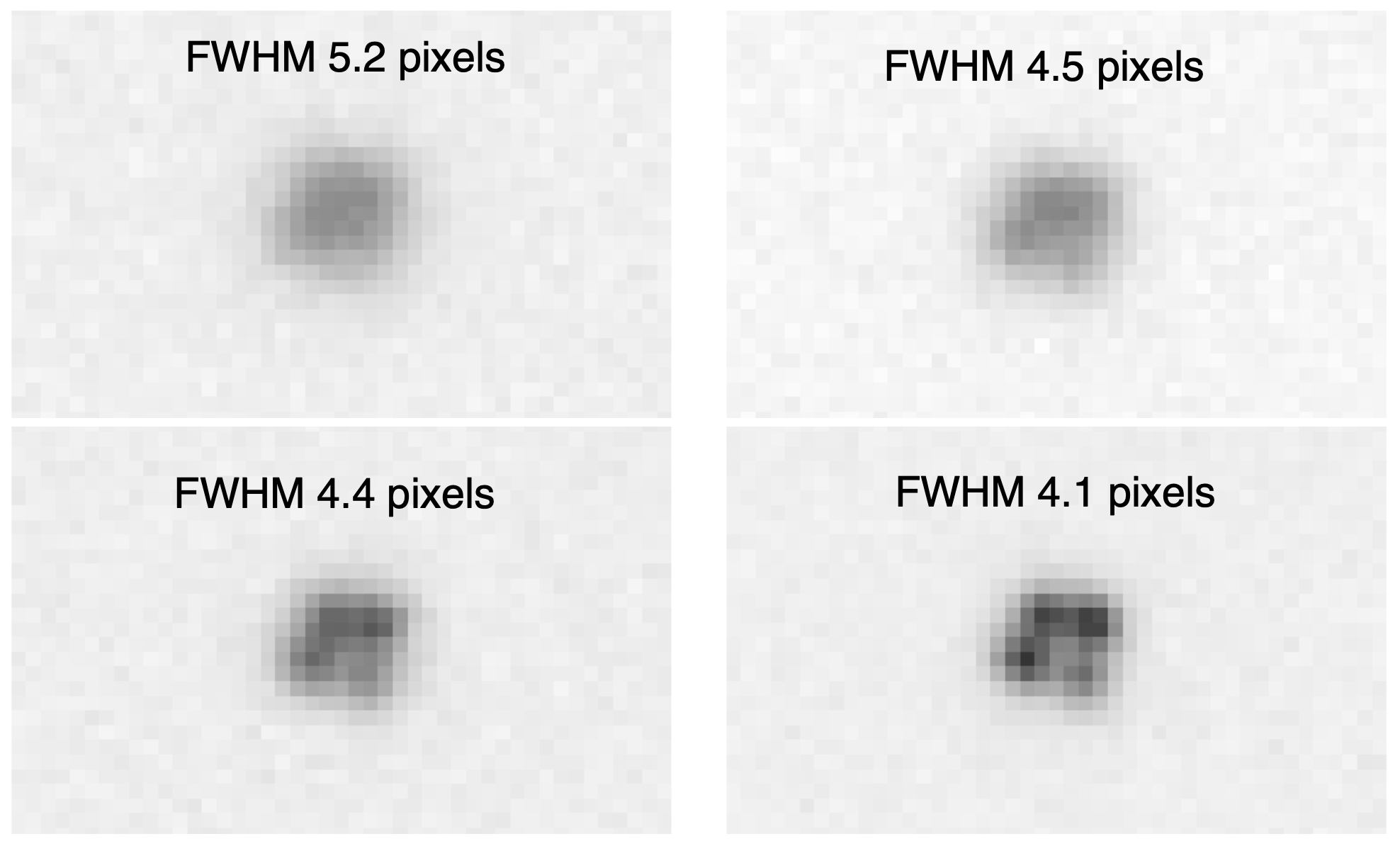}
  \caption{The impact of small differences in DIQ is very stark and
    motivates our recommendation to observe quad lenses in the latter
    stages of science verification (Section~\ref{sec:verify}). Here we
    show SDSS $r$-band images of DES\,J0405$-$3308 (a quadruply imaged
    quasar) from the DECam instrument on the Blanco telescope. Each
    cut-out is from an observation with an exposure time of 90
    seconds, and is based on real data, not simulations. The pixel
    scale is 0.263 arcsec, and thus the full width at half maximum
    (FWHM) of point sources in each panel is $\rm FWHM=1.37''$ (top
    left), $\rm FWHM=1.18''$ (top right), $\rm FWHM=1.16''$ (bottom
    left), and $\rm FWHM=1.08''$ (bottom right). All panels are scaled
    to the same stretch, and the photometric zeropoints are the same
    within 0.05 magnitudes.}
  \label{fig:quasars}
\end{figure}

We emphasize that the visual impression one gets of a lensed quasar is
exquisitely sensitive to small variations in image quality. In
Figure~\ref{fig:quasars} we show images of the quadruply lensed quasar
DES\,J0405$-$3308, taken with DECam on the Blanco
telescope. Relatively small differences in the DIQ produce dramatic
differences in its appearance. Visual inspection of the data at the
telescope during science verification observations will therefore give
the commissioning team immediate valuable feedback on DIQ variations
before the data are analyzed and quantitative measurements are made.

We recommend that each lens is observed through a single filter, ten
times during a visit: twice at the center of the field, and once at
each the corner of the 3 degree field, and once each at 3, 6, 9 and 12
o'clock. Repeat visits -- returning to a quad quasar to repeat this
series of ten observations -- will also permit assessment of DIQ and
feedback on the performance of the active optics system in different
seeing conditions and telescope state, and provide invaluable tests of
the reproducibility of test results.

In summary, we recommend to test the DIQ performance of the Simonyi
Telescope in the most demanding manner possible in the latter stages
of science verification. These tests are ideally accomplished through
the observations of the targets described here, supported by an
individual with strong lensing expertise embedded within Rubin’s
SIT-Com team. This individual would interact closely and rapidly with
hands-on commissioning colleagues. The core activity would be to
analyze Rubin data and measure covariances between observed quantities
that depend sensitively on image quality and system / atmospheric
quantities including DIMM seeing, airmass, windshake, mirror
temperature and residual wavefront error at varying distances from the
field center.

\section{Recommended fields for science validation surveys}
\label{sec:validate}

\noindent
During science validation it is essential to validate that the Rubin
Observatory and her Science Pipelines are capable of delivering
science-grade data on strong gravitational lenses. We stress that
before the ORR, we are interested in validating the robustness of the
pixel data and the catalogues derived from those pixels, and not on
science. Robust validation tests require commissioning observations of
fields that contain a large number of strong lenses that have been
discovered and studied by precursor surveys. These considerations, and
looking ahead to early science in DP2, both motivate prioritizing
equatorial fields for science validation surveys. This is because
observing to the North of Cerro Pach\'on maximizes the overlap with
both precursor surveys and observatories that can conduct follow-up
observations of new discoveries based on DP2.

We recommend specific wide fields and single pointings that satisfy
the over-arching criteria outlined above in Sections~\ref{sec:wide}
and \ref{sec:single} respectively. In summary the wide fields
prioritize equatorial CFHT-LS fields, and the single pointings include
carefully selected galaxy- and cluster-scale lenses, in addition to
the COSMOS and XMM-LSS Deep Drilling Fields. All recommended science
validation fields are listed in Table~1, including the months of the
year in which each field is observable for ${\ge}2$ hours per night at
$\rm airmass{\le}1.5$. We also summarise our preferred science
validation fields by quarter in Table~2. This table highlights that
there is overlap in our preferred fields with those listed in Table~2
of the DESC Commissioning Note (Amon et al., 2020).

\subsection{Wide fields for science validation}\label{sec:wide}

\noindent
We recommend that wide field science validation surveys observe the
equatorial fields from CFHT-LS, namely W1, W2, W4 (Table~1). These are
the most thoroughly explored regions of the sky for strong lens
discovery, both initially by colleagues using CFHT-LS data, and more
recently by colleagues in surveys (KIDS, HSC-SSP) whose footprints
expand on the original CFHT-LS fields. In anticipation of ten-year
depth observations being limited to ${\sim}100{-}200\,{\rm degree}^2$,
these CFHT-LS fields offer the perfect criterion on which to select a
sub-region of these wider and newer surveys. Should time allow, we
recommend to expand these CFHT-LS wide survey footprints in to the
surrounding KIDS and HSC-SSP equatorial fields.

We \emph{strongly} disfavor observing a ${\sim}100{-}200\,{\rm
  degree}^2$ region of the DES footprint during science validation
because the DES image quality and depth is generally not as good as
that of CFHT-LS and HSC-SSP. However, the CFHT-LS W4 field discussed
above overlaps with the SDSS Stripe 82, which is included within the
DES footprint. Therefore if the observing season allows, a
W4/S82-based science validation survey may be optimal (Table~2).

\subsection{Single pointings for science validation}\label{sec:single}

\noindent 
Among the LSST Deep Drilling Fields, we \emph{strongly} prefer XMM-LSS
and COSMOS because they match the LSSTCam field of view well, are
equatorial, and have been searched previously for strong lenses. Also,
the XMM-LSS field is located within the CFHT-LS W1 field discussed
above. Equally, we \emph{strongly} disfavor the ECDFS and ELAIS S1 as
science validation fields because they don't satisfy these criteria.

Turning to pointed observations of known strong lenses, we agree with
the galaxy-scale lenses put forward by DESC (Amon et al., 2020), and
include them in our Table~1. We prefer clusters that are (1) known as
strong lenses, (2) are well studied, (3) are at equatorial
declinations, and (4) that have larger Einstein radii. Applying these
criteria to the known cluster lenses, we obtain the clusters listed in
Table~1. They span the full range of right ascension, and many of them
are also listed by Amon et al. (2020). In a nuthsell, the optimal
cluster-scale lens for science validation observations is one of the
Hubble Frontier Fields, due to the wealth of superb data from
precursor surveys. Among the Frontier Fields, Abell\,370 is our
highest priority due to it being equatorial.

\section*{Acknowledgments}

\noindent
We thank our colleagues in the Strong Lensing Science Collaboration
and Dark Energy Science Collaboration Strong Lensing Topical Team for
their support and assistance during the preparation of this note. We
thank Keith Bechtol in particular for his patient support, advice, and
encouragement. GPS acknowledges support from The Royal Society,
Leverhulme Trust, and the Science and Technology Facilities Council.

\section*{References}

\noindent
Amon A., et al., 2020, arXiv:2010.15318\\
Collett T., Auger M., 2014, MNRAS, 443, 969\\
Jullo E., et al., 2010, Science, 329, 924\\
Smith G.\ P., et al., 2019, arXiv:1902.05140\\
Suyu S.\ H., et al., 2013, ApJ, 766, 70\\

\begin{table}[h]
  \centering {\normalsize Table~1: Strong Lensing Science
    Collaboration Recommended Rubin Commissioning Fields\medskip}
             {\small
    \begin{tabular}{lllll}
    \hline
    \noalign{\medskip}
    Target name & R.A., Dec.\,[J2000] & Observable & Priority & Comment\\
    \noalign{\medskip}
    \hline
    \noalign{\medskip}
    \multispan{5}{\dotfill\normalsize\sc Science verification: Quadruply-imaged quasars\dotfill}\\
    \noalign{\medskip}
    0029$-$3814 & 00:29:41 $-$38:14:26 & May--Dec & A &  \cr
    0214$-$2104 & 02:14:16 $-$21:05:35 & Jun--Jan & B & \cr
    0420$-$4037 & 04:20:47 $-$40:37:27 & Jul--Mar & A & \cr
    SDSS\,J0924$+$0219 & 09:24:56 $+$02:19:25 & Nov--May & A$+$ & Time delay $=2\rm days$; Chandra ToO\cr
    1131$-$4419 & 11:31:00 $-$44:19:59 & Dec--Jul & A & \cr
    PS\,1606$-$2333 & 16:06:00 $-$23:33:22 & Feb--Sep & A & Time delays ${\sim}10{-}45\rm days$\cr
    WFI\,2026$-$4536 & 20:26:10 $-$45:36:27 & Apr--Nov & A & Time delay ${\simeq}19\rm days$\cr
    2100$-$4452 & 21:00:15 $-$44:52:07 & Apr--Nov & B & \cr
    \noalign{\medskip}
    \hline
    \noalign{\medskip}
    \multispan{5}{\normalsize\dotfill\sc Science validation: Galaxy-scale lenses\dotfill}\\
    \noalign{\medskip}
    HE\,0230$-$2130 & 02:32:33 $-$21:17:26 & Jul--Jan & A & Time delay $=16\rm days$\cr
    DES\,0408$-$5354  & 04:08:22 $-$53:54:00 & Jul--Mar & B & Compound lens; Time delay $>40\rm days$\cr
    SDSS\,J0924$+$0219 & 09:24:56 $+$02:19:25 & Nov--May & A$+$ & See above\cr
    HSC\,J142449$-$005322 & 14:24:49 $-$00:53:22 & Feb--Aug & A & Compound lens\cr
    PS\,1606$-$2333 & 16:06:00 $-$23:33:22 & Feb--Sep & A & See above\cr
    WFI\,2026$-$4536 & 20:26:10 $-$45:36:27 & Apr--Nov & B & See above\cr
    \noalign{\medskip}
    \hline
    \noalign{\medskip}
    \multispan{5}{\normalsize\dotfill\sc Science validation: Deep drilling fields\dotfill}\\
    \noalign{\medskip}
    ELAIS S1 & 00:37:48 $-$44:00:00 & May--Dec & D & Not searched for strong lenses\cr
    XMM-LSS & 02:22:50 $-$04:45:00 & Jul--Jan & A$+$ & Overlaps CFHT-LS W1\cr
    ECDFS & 03:32:30 $-$28:06:00 & Jul--Feb & D & Not searched for strong lenses\cr
    COSMOS & 10:00:24 $+$02:10:55 & Dec--May & A & Overlaps CFHT-LS D2\cr
    \noalign{\medskip}
    \hline
    \noalign{\medskip}
    \multispan{5}{\dotfill\normalsize\sc Science validation: Galaxy cluster lenses\dotfill}\\
    \noalign{\medskip}
    Abell\,2744            & 00:14:21 $-$30:23:50 & May--Dec & B & Hubble Frontier Field\\
    MACS\,J0138.0$-$2155   & 01:38:04 $-$21:55:49 & Jun--Jan & A & \cr
    Abell\,370             & 02:39:53 $-$01:34:37 &  Jul--Jan & A$+$ & Hubble Frontier Field\\
    MACS\,J0416.1$-$2403   & 04:16:09 $-$24:04:29 & Jul--Mar & A$+$ & Hubble Frontier Field\\
    RXC\,J0600.1$-$2007    & 06:00:10 $-$20:08:09 & Aug--Apr & A & \cr
    1E0657$-$558             & 06:58:38 $-$55:57:00 & Sep--May & B & Bullet cluster\cr
    Abell\,0868            & 09:45:26 $-$08:39:06 & Nov--Jun & A & \cr
    MACS\,J1206.2$-$0847   & 12:06:12 $-$08:48:02 & Dec--Jul & A & \cr
    Abell\,1689            & 13:11:34 $-$01:21:56 & Jan--Jul & A & \cr
    Abell\,1835            & 14:01:02 $+$02:52:43 & Jan--Jul & A & \cr
    Abell\,2204              & 16:32:47 $+$05:34:14 & Mar--Aug & A & \cr
    PLCK\,G004.5$-$19.5    & 19:17:05 $-$33:31:29 & Mar--Oct & B & \cr
    MACS\,J2140.2$-$2339   & 21:40:15.2 $-$23:39:40 & Apr--Nov & A & Bright radial arc\cr
    Abell~S1063            & 22:48:44 $-$44:31:49 & May--Dec & B & Hubble Frontier Field\\
    \noalign{\medskip}
    \hline
    \noalign{\medskip}
    \multispan{5}{\normalsize\dotfill\sc Science validation: Wide survey fields\dotfill}\\
    \noalign{\medskip}
    CFHT-LS W1 & 02:18:00 $-$07:00:00 & Jul--Jan & A & Overlaps XMM-LSS\cr
    CFHT-LS W2 & 08:57:49 $-$03:19:00 & Nov--May & A & \cr
    CFHT-LS W4 & 22:13:18 $+$01:19:00 & May--Nov & A$+$& Overlaps DES, KIDS, HSC, S82\cr
    \noalign{\medskip}
    \hline
    \end{tabular}
    }
\end{table}

\begin{table}[ht]
  \centering
    {\normalsize Table~2: Strong Lensing Science Collaboration recommended science validation survey scenarios by quarter\medskip}
    {\small
    \begin{tabular}{p{20mm}p{35mm}p{35mm}p{60mm}}
    \hline
    \noalign{\medskip}
    Field type & Preferred & Back-up & Comment \\
    \noalign{\medskip}
    \hline
    \noalign{\medskip}
    \multispan{4}{\normalsize\dotfill\sc January--March\dotfill}\\
    \noalign{\medskip}
    Wide-field & CFHT-LS W2 & CFHT-LS W1 & \cr
    Galaxy lens & SDSS\,J0924$+$0219 & DES\,0408$-$5354 & \cr
    Cluster lens & A\,0868 & MACS\,J1206.2$-$0847 & \cr
    LSST DDF & COSMOS & ECDFS & Strong preference {\bf against} ECDFS. \cr
    \noalign{\medskip}
    \hline
    \noalign{\medskip}
    \multispan{4}{\normalsize\dotfill\sc April--June\dotfill}\\
    \noalign{\medskip}
    Wide-field & CFHT-LS W2 & CFHT-LS W4 & \cr
    Galaxy lens & SDSS\,J0924$+$0219 & HSC\,J142449$-$005322 & \cr
    Cluster lens & Abell\,1689 & Abell\,2204 & \cr
    LSST DDF & COSMOS & ELAIS S1 & Strong preference {\bf against} ELAIS S1. \cr
    \noalign{\medskip}
    \hline
    \noalign{\medskip}
    \multispan{4}{\normalsize\dotfill\sc July--September\dotfill}\\
    \noalign{\medskip}
    Wide-field & CFHT-LS W4 & CFHT-LS W1 & \cr
    Galaxy lens & HE\,0230$-$2130 & PS\,1606$-$2333 & \cr
    Cluster lens & MACS\,J2140.2$-$2339 & Abell\,2744 & Abell\,2744 is a Hubble Frontier Field. \cr
    LSST DDF & XMM-LSS & ELAIS S1 & Strong preference {\bf against} ELAIS S1. \cr
    \noalign{\medskip}
    \hline
    \noalign{\medskip}
    \multispan{4}{\normalsize\dotfill\sc October--December\dotfill}\\
    \noalign{\medskip}
    Wide-field & CFHT-LS W4 & CFHT-LS W1 & \cr
    Galaxy lens & SDSS\,J0924$+$0219 & HE\,0230$-$2130 & \cr
    Cluster lens & MACS\,J0416.1$-$2403 & Abell\,370 & Both are Hubble Frontier Fields. \cr
    LSST DDF & XMM-LSS & ECDFS & Strong preference {\bf against} ECDFS. \cr
    \noalign{\medskip}
    \hline
    \end{tabular}
    }
\end{table}

\end{document}